\documentclass[12pt]{article}
\usepackage{amssymb,amsmath}
\usepackage[noblocks]{authblk}
\usepackage[top=0.75in, bottom=0.75in, left=0.75in, right=0.75in, dvips]{geometry}
\usepackage{caption}
\date{}

\begin{document}
\textwidth 10.0in   
\textheight 9.0in 
\topmargin -0.60in
\title{Supersymmetric Models on $AdS_3$ and $AdS_4$ Embedding Superspaces}
\author[1]{D.G.C. McKeon}
\affil[1] {Department of Applied Mathematics, The
University of Western Ontario, London, ON N6A 5B7, Canada}

\maketitle

\maketitle

\begin{abstract}
Superspace techniques are used to formulate a supersymmetric model on an $AdS_3$ surface embedded in four dimensions.  In this model, the supersymmetry transformation is the ``square root'' of transformation generated by the isometry generators of $AdS_3$.  Since momentum is not an isometry generator, supersymmetry does not result in equal masses for a Bosonic field and its Fermionic partner.  We express this model in terms of coordinates that characterize the $AdS_3$ space. In one coordinate system, it is possible to define a subspace with a Minkowski metric. It becomes possible to infer a model in $AdS_4$ space in which there is a symmetry transformation that relates Bosonic and Fermionic fields.  This model is not a consequence of being formulated in a superspace and the Fermionic  symmetry transformation is not the ``square root'' of an isometry of $AdS_4$.  
\end{abstract}

\section{Introduction}
Much work has been done on formulating field theory on spaces of constant curvature (see, for example [1-13]).  This topic has become more prominent with the introduction of the AdS/CFT conjecture [14, 15]. The formulation of supersymmetric models on AdS space has generally been a consequence of more general supergravity models. However, it is also possible to formulate supersymmetric models on spaces of constant curvature by considering these spaces to be embedded in spaces of higher dimension [16, 17].

In this paper we extend the work done on supersymmetric models in $AdS_3$ and $AdS_4$ presented in ref. [17].  In the embedding space for $AdS$, it proves possible to use superspace techniques to formulate a model possessing a symmetry whose algebra is that of the supersymmetric extension of the $AdS_3$ isometry algebra.  This model can be expressed in terms of coordinates that lie in the $AdS_3$ space.

It then proves possible to formulate a model for $AdS_4$ that possesses a Fermionic symmetry using the embedding space coordinates.  Although this model is analogous to the supersymmetric model in $AdS_3$, it turns out that the symmetry which relates Bosons to Fermions in this model does not have an algebra that is related to the isometry algebra for $AdS_4$. 

In both $AdS_3$ and $AdS_4$, the masses of the Fermionic and Bosonic fields can be distinct as the supersymmetry generator is not the ``square root'' of a momentum generator.

\section{Supersymmetry in $AdS_3$}
The space $AdS_3$ can be viewed as being the three dimenional surface embedded in a four dimensional space $x^A$ $(A = 1, \ldots , 4)$ defined by the equation
\begin{equation}
\eta_{AB} x^Ax^B = r^2
\end{equation}
where $\eta_{AB}$ is the metric $\eta_{AB} = \rm{diag} (+1, +1, -1, -1)$. 

The generators of an isometry transformation on this surface are $M_{AB}$; they satisfy the algebra
\begin{equation}
[M_{AB}, M_{CD} ] = \eta_{AC} M_{BD} - \eta_{AD} M_{BC} +\eta_{BD}M_{AC} - \eta_{BC} M_{AD}.
\end{equation}
This algebra is realized by the angular momentum operator
\begin{equation}
L_{AB} = -x_A \partial_B + x_B \partial_A
\end{equation}
as well as the two spin operators $\sigma_{AB}$ and $\overline{\sigma}_{AB}$ where (with $\epsilon^{1234} = 1$)
\begin{subequations}
\begin{align}
\sigma_{AB} = -\frac{1}{4} (\lambda_A \overline{\lambda}_B - \lambda_B \overline{\lambda}_A) = - \frac{1}{2}\epsilon_{ABCD} \sigma^{CD}\\
\intertext{and}\nonumber \\
\overline{\sigma}_{AB} = -\frac{1}{4} (\overline{\lambda}_A \lambda_B - \overline{\lambda}_B \lambda_A) = + \frac{1}{2}\epsilon_{ABCD}\, \overline{\sigma}^{CD}.
\end{align}
\end{subequations}
In eq. (4), $\lambda_A$ and $\overline{\lambda}_A$ are $2 \times 2$ matrices related to the usual Pauli spin matrices $\vec{\tau}$ by
\begin{subequations}
\begin{align}
\lambda_A  = (1, i\tau^2, \tau^1 , \tau^3)\\
\intertext{and}\nonumber \\
\overline{\lambda}_A = (1, -i\tau^2, -\tau^1 , -\tau^3).
\end{align}
\end{subequations}
Since
\begin{equation}
\lambda^A \overline{\lambda}^B + \lambda^B \overline{\lambda}^A = 2\eta^{AB}
\end{equation}
the Dirac matrices
\begin{equation}
\gamma^A = \left( \begin{array}{cc}
0 & \lambda^A \\
\overline{\lambda}^A & 0 \end{array} \right)
\end{equation}
satisfy
\begin{equation}
\left\lbrace \gamma^A, \gamma^B \right\rbrace = 2 \eta^{AB}.
\end{equation}
We have occasion to make use of the Fierz identities
\begin{subequations}
\begin{align}
(\lambda^A)_{ij} (\overline{\lambda}_A)_{k\ell} = 2 \delta_{i\ell} \delta_{kj} \\
\intertext{and}\nonumber \\
2 \lambda_{ij}^A \delta_{k\ell} = \lambda_{i\ell}^A \delta_{kj} + \epsilon^{ABCD} \lambda_{Bi\ell} \sigma_{CD\,kj} \\
2 \overline{\lambda}_{ij}^A \delta_{k\ell} = \overline{\lambda}_{i\ell}^A \delta_{kj} - \epsilon^{ABCD} \overline{\lambda}_{Bi\ell} \overline{\sigma}_{CD\,kj}.
\end{align}
\end{subequations}
In addition we find that
\begin{subequations}
\begin{align}
\sigma^{AB }\lambda^C &- \lambda^C \overline{\sigma}^{AB} = \eta^{AC}\lambda^B - \eta^{BC}\lambda^A\\
\lambda^A \overline{\lambda}^B \lambda^C &= \eta^{AB} \lambda^C - \eta^{AC} \lambda^B + \eta^{BC} \lambda^A + \epsilon^{ABCD} \lambda_D . 
\end{align}
\end{subequations}

We also note that since
\begin{subequations}
\begin{align}
\tau^2 \lambda^A \tau^2 &= \overline{\lambda}^{AT} \\
\tau^2 \sigma^{AB} \tau^2 &= -\sigma^{ABT}
\end{align}
\end{subequations}
we can define the matrix
\begin{equation}
C = i\lambda^3 \lambda^4 = \left( \begin{array}{cc}
-\tau^2 & 0 \\
0 & -\tau^2 \end{array} \right)
\end{equation}
so that
\begin{equation}
C \gamma^\mu C^{-1} = \gamma^{\mu\dagger} = \gamma^{\mu T}.
\end{equation}
The charge conjugate of a four component Dirac spinor $\Psi$ is given by 
\begin{equation}
\Psi_C = C \overline{\Psi}^T
\end{equation}
where
\begin{equation}
\overline{\Psi} = - \Psi^\dagger C; 
\end{equation}
since $(\Psi_C)_C = \Psi$, a Dirac spinor can be taken to be simultaneously Weyl and Majorana [18].  This means that if
\begin{equation}
\Psi = \left( \begin{array}{c} 
\theta \\
\phi 
\end{array} \right)
\end{equation}
then the two component spinors $\theta$ and $\phi$ are both real.  Furthermore, under the transformation
\begin{equation}
\Phi \longrightarrow e^{\omega^{AB}\sum_{AB}} \Psi
\end{equation}
where $\sum_{AB} = - \frac{1}{4}\left[\gamma_A, \gamma_B \right]$, then 
\begin{subequations}
\begin{align}
\theta &\longrightarrow e^{\omega^{AB}\sigma_{AB}}\theta \\
\tilde{\theta} &\longrightarrow \tilde{\theta} e^{-\omega^{AB}\sigma_{AB}} \\
\phi &\longrightarrow e^{\omega^{AB}\overline{\sigma}_{AB}}\phi \\
\tilde{\phi} &\longrightarrow \tilde{\phi} e^{-\omega^{AB}\overline{\sigma}_{AB}}
\end{align}
\end{subequations}
where
\begin{equation}
\tilde{\theta} = \theta^T\tau^2, \quad \tilde{\phi} = \phi^T \tau^2. 
\end{equation}
Under the transformation of eq. (18), $\tilde{\theta}\theta$ and $\tilde{\phi}\phi$ are scalars, $\tilde{\phi}\overline{\lambda}^A\theta$ and $\tilde{\theta}\lambda^A \phi$ are vectors (by eq. (10a)) while $\tilde{\theta}\sigma^{AB}\theta$ and $\tilde{\phi}\overline{\sigma}^{AB}\phi$ are tensors (by eq. (2)).  However, if $\Psi$ is Grassmann, then $\tilde{\theta}\sigma^{AB}\theta = \tilde{\phi}\overline{\sigma}^{AB}\phi = 0$. 

For functional differentiation with respect to the spinor $\theta$ it is apparent that 
\begin{equation}
\frac{\partial}{\partial\theta_i} = \frac{\partial\tilde{\theta}_j}{\partial\theta_i} \frac{\partial}{\partial\tilde{\theta}_j} = \tau_{ij}^2 \frac{\partial}{\partial\tilde{\theta}_j}
\end{equation}
so that by eqs. (11b, 18) we have the transformations
\begin{equation}
\frac{\partial}{\partial\tilde{\theta}} \longrightarrow e^{\omega^{AB}\sigma_{AB}} \frac{\partial}{\partial\tilde{\theta}}.
\end{equation}
If we take a generator $Q$ to be a Majorana-Weyl spinor in $2 + 2$ dimensions, then the Fierz identity for $\sum_{AB}$
\begin{equation}
({\textstyle\sum^{AB}})_{ij} ({\textstyle\sum_{AB}})_{k\ell} = -\frac{1}{2} ({\textstyle\sum^{AB}})_{i\ell} ({\textstyle\sum_{AB}})_{kj} - \frac{3}{4} \left(\delta_{i\ell} \delta_{kj} + \gamma_{i\ell}^5\gamma_{kj}^5 \right)
\end{equation}
$(\gamma^5 = - \gamma^0\gamma^1\gamma^2\gamma^3)$ can be used to show that the supersymmetry algebra
\begin{subequations}
\begin{align}
\left\lbrace Q, \tilde{Q}\right\rbrace &= 2\, {\textstyle\sum^{AB}} M_{AB} \\
\left[ M^{AB}, Q \right] &= - {\textstyle\sum^{AB}} Q
\end{align}
\end{subequations}
satisfies the Jacobi identity.  If now we take
\begin{subequations}
\begin{align}
K^{AB} &= \frac{1}{2}\left(M^{AB} - \frac{1}{2} \epsilon^{ABCD} M_{CD}\right) = - \frac{1}{2} \epsilon^{ABCD} K_{CD} \\
\overline{K}^{AB} &= \frac{1}{2}\left(M^{AB} + \frac{1}{2} \epsilon^{ABCD} M_{CD}\right) = + \frac{1}{2} \epsilon^{ABCD} \overline{K}_{CD}
\end{align}
\end{subequations}
\begin{subequations}
\begin{align}
Q &= \lambda^A \left( \theta \partial_A + x_A \frac{\partial}{\partial\tilde{\theta}}\right) \\
\tilde{Q} &=  \left( \tilde{\theta} \partial_A - \frac{\partial}{\partial\theta} x_A \right)\overline{\lambda}^A
\end{align}
\end{subequations}
\begin{subequations}
\begin{align}
R &= \overline{\lambda}^A \left( \theta \partial_A + x_A \frac{\partial}{\partial\tilde{\theta}}\right) \\
\tilde{R} &=  \left( \tilde{\theta}\partial_\lambda - \frac{\partial}{\partial\theta} x_A \right)\lambda^A
\end{align}
\end{subequations}
then the algebra of eq. (23) decomposes into
\begin{subequations}
\begin{align}
\left\lbrace Q,\tilde{Q} \right\rbrace &= 2\sigma^{AB} K_{AB} \\
\left[ K^{AB}, Q \right] & = - \sigma^{AB} Q 
\end{align}
\end{subequations}
and
\begin{subequations}
\begin{align}
\left\lbrace R,\tilde{R} \right\rbrace &= 2\overline{\sigma}^{AB} \overline{K}_{AB} \\
\left[ \overline{K}^{AB}, Q \right] & = - \overline{\sigma}^{AB} Q , 
\end{align}
\end{subequations}
with $K^{AB}$ and $\overline{K}^{AB}$ both satisfying eq. (2) (ie, we can make the decomposition $SO(2,2) = SO(2,1) \times SO(2,1)$).

We now introduce a scalar superfield $\Phi(x^A, \theta)$
\begin{equation}
\Phi (x^A, \theta) = \phi (x^A) + \tilde{\theta}\rho (x^A) + \frac{1}{2} F(x^A) \tilde{\theta}\theta .
\end{equation}
With $R$ being given by eq. (26a), we see that this induces a change $\delta\Phi = [\tilde{\epsilon}R, \Phi ]$, which, upon using eq. (9b), results in
\begin{subequations}
\begin{align}
\delta\phi &= \tilde{\epsilon}\overline{\lambda} \cdot x \rho = \tilde{\rho} \lambda \cdot x \epsilon \\
\delta\rho &= \lambda^A (\phi_{,A} + F x_A)\epsilon \\
\delta F &= - \tilde{\epsilon}\overline{\lambda}^A \rho_{,A} = - \tilde{\rho}_{,A} \lambda^A\epsilon\,.
\end{align}
\end{subequations}
(We have $\tilde{\epsilon}$ transforming as $\tilde{\phi}$ does in eq. (18d).)  It can be verified directly that the transformations of eq. (30) are consistent with eq. (28).  We also find that 
\begin{equation}
\left[ \tilde{\epsilon} R, x^Ax_A - \tilde{\theta}\theta\right] = 0.
\end{equation}

Functional integration over the Grassmann variable $\theta$ is normalized so that
\begin{subequations}
\begin{align}
\int d^2 \theta \,\tilde{\theta}\theta &= 1 \\
\intertext {or equivalently}\nonumber \\
\int d^2 \theta \,\theta_i\theta_j &= - \frac{1}{2} \tau_{ij}^2.
\end{align}
\end{subequations}
The product of superfields is clearly a superfield and by eq. (30c) its $\tilde{\theta}\theta$ component is a total derivative.  This observation and eq. (31) lead us to consider the action
\begin{equation}
S = \int d^4x\, d^2\theta \,\delta \left(x^A x_A - \tilde{\theta}\theta - r^2\right) \left[ \Phi \tilde{R}R \Phi + g_2 \Phi^2\right]
\end{equation}
as it is invariant by construction under the transformation of eq. (30).  (Interaction terms $g_n\Phi^n$ could also be included.) 

We now use the relation
\begin{align}
\int d^4x d^2\theta f \,&\delta \left( x^Ax_A - \tilde{\theta}\theta - r^2\right) \nonumber \\
& = \int d^4x d^2\theta f(x) \left[ \delta (x^A x_A - r^2) - \tilde{\theta}\theta \delta^\prime(x^Ax^A - r^2)\right] \nonumber \\
& = \int d^4x d^2\theta \delta(x^Ax_A - r^2) \left[ f(x) + \tilde{\theta}\theta \partial_A \left(\frac{x^A}{2x^Bx_B} f(x)\right)\right]
\end{align}
(since $x^B \partial_B \delta (x^Ax_A - r^2) = 2x^B x_B \delta^\prime(x^A x_A - r^2))$ as well as
\begin{align}
\tilde{R}R &=  \tilde{\theta}\theta \partial^A \partial_A + \tilde{\theta}\frac{\partial}{\partial \tilde{\theta}} \left( 4 + 2x^A\partial_A - 2\overline{\sigma}^{AB} L_{AB}\right) \nonumber \\
&- 2x^A\partial_A - x^A x_A \frac{\partial}{\partial\theta} \frac{\partial}{\partial\tilde{\theta}}
\end{align}
to rewrite $S$ in eq. (33) as
\begin{align}
S = \int d^4x & \delta(x^2 - r^2) \big[ \frac{1}{r^2} \phi \left( \frac{1}{2} L^{AB} L_{AB} - \omega^2 + g_2 (\omega + 1)\right) \phi \nonumber \\
&+ \tilde{\rho} \left( \overline{\sigma}^{AB}L_{AB} -2 - \frac{1}{2} g_2\right)\rho - \frac{r^2}{2} F^2 + (1 - \omega + g_2) \phi F\big] .
\end{align}
In eq. (36) we have used the $d$-dimensional relation
\begin{equation}
\partial^A \partial_A  = \frac{1}{2x^Ax_A}
\left[ L^{AB} L_{AB} + 2 (d - 2) x^A \partial_A + 2 (x^A \partial_A)^2\right]
\end{equation}
and have assumed that
\begin{equation}
\Delta \Phi \equiv \left( x^A\partial_A + \tilde{\theta} \frac{\partial}{\partial \tilde{\theta}}\right) \Phi = \omega \Phi .
\end{equation}
This last relationship is a supersymmetric generalization of the homogeneity condition of Dirac [1] and is consistent with supersymmetry as $[\tilde{\epsilon} R, \Delta] = 0$. 

We note from eq. (36) that the field $F$ is an auxiliary field, and that the scalar $\phi$ and the spinor $\rho$ have distinct masses.  Supersymmetric models in Minkowski space have equal masses for such pairs of fields as the symmetry generator is the ``square root'' of the momentum generator [19].

We now consider how the action of eq. (36) can be expressed in terms of coordinates on the curved surface defined by eq. (1) rather than the coordinates $x^A$ of the embedding space.  If we take
\begin{equation}
x^A = r f^A(y^a) \qquad (a = 1,2,3)
\end{equation}
where $\eta_{AB} f^Af^B = 1$, then the line element in the embedding space is given by
\begin{align}
ds^2 &= \eta_{AB} dx^A dx^B \nonumber \\
&= dr^2 + \eta_{AB}r^2 \frac{\partial f^A}{\partial y^a}
\frac{\partial f^B}{\partial y^b} dy^a dy^b
\end{align}
so that we can define the metric 
\begin{equation}
g_{ab} =  \eta_{AB} \frac{\partial f^A}{\partial y^a}
\frac{\partial f^B}{\partial y^b} .
\end{equation}
The operator $L_{AB}$ of eq. (2) is evidently a linear combination of the derivatives $\frac{\partial}{\partial y^a}$ and so
\begin{equation}
L^{AB} = \Lambda^{ABa} \frac{\partial}{\partial y^a} = -rf^A\frac{\partial}{\partial x_B} + rf^B \frac{\partial}{\partial x_A }.
\end{equation}
As $\frac{\partial}{\partial y^a} = r \frac{\partial f^A}{\partial y^a} \frac{\partial}{\partial x^A}$, eq. (42) leads to 
\begin{equation}
\Lambda^{ABa} \frac{\partial f^C}{\partial y^a} = - f^A \eta^{BC} + f^B \eta^{AC} .
\end{equation}
Together, eqs. (41) and (43) show that 
\begin{equation}
\Lambda^{ABa} = g^{ab} \left( -f^A \frac{\partial f^B}{\partial y^b} + f^B \frac{\partial f^A}{\partial y^b}\right)
\end{equation}
and so 
\begin{align}
L_{AB} L^{AB} = 2\bigg[ \frac{\partial g^{cd}}{\partial y^d}
\frac{\partial}{\partial y^c} &+  g^{ab} g^{cd} \frac{\partial^2 f^A}{\partial y^a\partial y^d} \frac{\partial f_A}{\partial y^b} \frac{\partial}{\partial y^c} \\
&+ g^{ab} \frac{\partial^2}{\partial y^a \partial y^b}\bigg]. \nonumber
\end{align}
If now $g = det\, g_{ab}$, then we obtain
\begin{equation}
\frac{\partial g}{\partial y^a} = \frac{\partial g}{\partial g_{mn}} \frac{\partial g_{mn}}{\partial y^a} = g\,g^{mn} \frac{\partial g_{mn}}{\partial y^a} .
\end{equation}
Together, eqs. (45, 46) lead to 
\begin{equation}
\frac{1}{2} L^{AB} L_{AB} = \frac{1}{\sqrt{g}} \frac{\partial}{\partial y^a} \left(g^{ab} \sqrt{g} \right) \frac{\partial}{\partial y^a}.
\end{equation}

On the surface defined by eq. (1) we first take
\begin{equation}
(x^1, x^2, x^3, x^4) = (r \sec \rho \sin t, \;\;  r \sec \rho \cos t , \;\; r \tan \rho \sin \psi ,\;\; r \tan \rho \cos \psi )
\end{equation}
with $r$ being a constant. In general, with $ds^2$ being given by eq. (40),
\begin{equation}
ds^2 = dr^2 + r^2 \left[ \sec^2 \rho (dt^2 - d\rho^2) - \tan^2\rho\, d\psi^2 \right]
\end{equation}
and by eq. (47)
\begin{equation}
\frac{1}{2} L^{AB} L_{AB} = cos^2\rho \left( \frac{\partial^2}{\partial t^2} - \frac{\partial^2}{\partial \rho^2}\right) - \cot \rho \frac{\partial}{\partial \rho} - \cot^2 \rho \frac{\partial^2}{\partial \rho^2} .
\end{equation}
Furthermore, with these coordinates, eqs. (4b, 5, 44) lead to 
\begin{align}
\overline{\sigma}^{AB} L_{AB} = i \tau^2 & (\partial_\psi - \partial_t) + \left[\sin (t - \psi) - i \tau^2 \cos(t -\psi)\right]\\
& \left[ (\sin \rho \partial_t - \csc\rho \partial_\psi)\tau^1 - \cos \rho \partial_\rho \tau^3\right]. \nonumber
\end{align}
If now we set
\begin{equation}
\tan\rho = \sinh \kappa
\end{equation}
then eqs. (50) and (51) become
\begin{subequations}
\begin{align}
\frac{1}{2} L^{AB} L_{AB} &= \textrm{sech}^2\kappa \frac{\partial^2}{\partial t^2} - \frac{\partial^2}{\partial \kappa^2} - (\coth \kappa \tanh \kappa ) \frac{\partial}{\partial\kappa} - \cosh^2\kappa \frac{\partial^2}{\partial\psi^2}\\
\intertext{and}
\overline{\sigma}^{AB} L_{AB} &= i\tau^2 (\partial_\psi - \partial_t) + \left[ \sin (t - \psi) - i \tau^2 \cos (t-\psi)\right] \\
&\hspace{3cm}\left[ (\tanh \kappa \partial_t - \coth \kappa \partial_\psi) \tau^1 - \partial_\kappa \tau^3 \right]\nonumber
\end{align}
\end{subequations}
respectively.

We can also use the Poincar$\acute{\textrm{e}}$ coordinates
\begin{subequations}
\begin{align}
r &= (x^{1^{2}} + x^{2^{2}} - x^{3^{2}} - x^{4^{2}})^{1/2} \\
a &= \frac{x^1 - x^4}{r^2}\\
t &= \frac{x^2}{ar}\\
z &= \frac{x^3}{ar} .
\end{align}
\end{subequations}
With these coordinates, the element of arc length in eq. (40) becomes
\begin{align}
ds^2 = - \frac{r^2}{a^2} da^2& + a^2r^2 (dt^2 - dz^2) + a^2 (t^2 - z^2)dr^2 \nonumber \\
 & - 2a^2 r\,dr \left( \frac{1}{a^3} da + t\, dt - z\,dz\right);
\end{align}
in addition we find that
\begin{equation}
x^A \partial_A = r \partial_r - a \partial_a + t\partial_t + z\partial_z.
\end{equation}
Furthermore, direct calculation shows that
\begin{equation}
\frac{1}{2} L^{AB} L_{AB} = - a^2 \partial_a^2 - 3a\partial_a + \frac{1}{a^2} \left(\partial_t^2 - \partial_z^2 \right)
\end{equation}
and
\begin{align}
\overline{\sigma}^{AB} L_{AB} = \frac{1}{2} &(\tau^1 - i\tau^2) \bigg[ \frac{a}{r} (t - z) \partial_a - \frac{1}{2a^2r} (\partial_t + \partial_z)\\
& + \frac{zt}{r} (\partial_t - \partial_z) + \frac{z^2 -t^2}{2r} (-\partial_t + \partial_z)\bigg]\nonumber \\
&+ \frac{r}{4} (\tau^1 + i\tau^2) (\partial_t - \partial_z) + \frac{\tau^3}{2} \left[ a \partial_a - (t -z) (\partial_t - \partial_z) \right].\nonumber
\end{align}
If we were to restrict our attention to the surface $r = \rm{const.}$, $a = \rm{const.}$ so that $dr = da = 0$, then by eq. (55) $ds^2 = (ar)^2 (dt^2 - dz^2)$ which is an element of arc length conformally equivalent to two dimensional Minkowski space.  If we take all fields to be independent of $r$ then by eqs. (38) and (56)
\begin{subequations}
\begin{align}
a\partial_a \phi &= (t \partial_t + z \partial_z - \omega)\phi\\
a\partial_a \rho &= \left[t \partial_t + z \partial_z - (\omega - 1)\right]\rho\\
\intertext{and}
a\partial_a F &= \left[ t \partial_t + z \partial_z - (\omega -2)\right]F .
\end{align}
\end{subequations}
(We note that $L_{34} = a\partial_a - t\partial_t - z \partial_z$ by eqs. (3) and (54).)  Eqs. (57) and (58) can be used in conjunction with eq. (59) to reduce the action of eq. (36) to a supersymmetric model in the two dimensional space spanned by the coordinates $t$ and $z$.  The supersymmetry is that of eq. (30), again provided $\phi$, $\rho$ and $F$ are independent of $r$ and their dependence on $a$ is dictated by eq. (59).  From eq. (58) it is not apparent if this model is invariant under a Poincar$\acute{\textrm{e}}$ transformation in the space spanned by $t$ and $z$.  It is however invariant under the Bosonic symmetry generated by $L_{AB}$; in particular by $L_{12} = - (t \partial_x + x \partial_t)$.

\section{Supersymmetry in $AdS_4$}

Supersymmetry using the five dimensional embedding space for $AdS_4$ has been discussed in refs. [16, 17].  This embedding space is defined by eq. (1) where now $A = (0, 1, 2, 3, 5)$ and $\eta_{AB} = \rm{diag} (+1, -1, -1, -1, +1)$.  A suitable set of Dirac matrices is
\begin{equation}
\Gamma^0 = \left(\begin{array}{cc}
0 & 1 \\
1 & 0 \end{array}\right) \quad
\Gamma^i = \left(\begin{array}{cc}
0 & \tau^i \\
-\tau^i & 0 \end{array}\right)  \quad
\Gamma^5 = \left(\begin{array}{cc}
-1 & 0 \\
0 & 1 \end{array}\right) 
\end{equation}
with the charge conjugation matrix $C$ given by 
\begin{equation}
C = i \Gamma^1\Gamma^3 = C^{-1} = C^\dagger = -C^* = - C^T.
\end{equation}
We let
\begin{equation}
\overline{\phi} = Q^\dagger (-i \Gamma^0\Gamma^5)
\end{equation}
for a spinor $\phi$ in the embedding space; if
\begin{equation}
\phi_C = C \overline{\phi}^T
\end{equation}
then $\phi = (\phi_c)_c$ and so a spinor can be taken to be Majorana. The supersymmetry of eqs. (2, 23) can now be shown to be consistent in this five dimensional embedding space associated with $AdS_4$ [17].  However, despite the fact that the supersymmetry algebra for $AdS_4$ can be represented in a superspace, it doesn't turn out to be feasible to define a supersymmetric model using this superspace [17].  We can, however, use the $AdS_3$ model of eq. (36) to find an analogous model defined in the embedding space for $AdS_4$ that contains a Fermionic symmetry.

An action on $AdS_4$ that is modelled on the action of eq. (36) is
\begin{equation}
S = \int d^5 x \delta (x^2 - r^2) \left[ \tilde{\Psi} \left( \textstyle{\sum}^{\mu\nu} L_{\mu\nu} + A\right) \Psi + B\Phi\left(L^{\mu\nu} L_{\mu\nu} + b\right)\Phi + \vert{\!\!\!C} F^2\right].
\end{equation}
In eq. (64) $\Psi$ is a Majorana spinor with $\tilde{\Psi} = \Psi^TC$, $\Phi$ and $F$ real scalars, and $A, B, \vert{\!\!\!C}$ and $b$ undetermined constants.  These constants are to be fixed by requiring that the action of eq. (64) be invariant under a Fermionic symmetry transformation of the form
\begin{subequations}
\begin{align}
\delta\Psi &= \left[\left(\textstyle{\sum}^{\mu\nu} L_{\mu\nu} + \alpha\right) \Phi - F\right]\xi\\
\delta\Phi &= \tilde{\xi} \Psi\\
\delta F & = -\tilde{\xi} \left(\textstyle{\sum}^{\mu\nu} L_{\mu\nu} + \gamma\right) \Psi 
\end{align}
\end{subequations}
where $\xi$ is a constant Majorana spinor and $\alpha$ and $\gamma$ are constants.

Substitution of eq. (65) into eq. (64),along with the useful equations
\begin{subequations}
\begin{align}
C \gamma^\mu C^{-1} &= \gamma^{\mu T}\\
C \textstyle{\sum}^{\mu\nu} C^{-1} &= -\textstyle{\sum}^{\mu\nu T}
\end{align}
\end{subequations}
\begin{equation}
\left[ \textstyle{\sum}^{\mu\nu}, \textstyle{\sum}^{\lambda\sigma}\right] = 
\eta^{\mu\lambda} \textstyle{\sum}^{\nu\sigma} - \eta^{\nu\lambda} \textstyle{\sum}^{\mu\sigma} + \eta^{\nu\sigma}\textstyle{\sum}^{\mu\lambda} - \eta^{\mu\sigma}\textstyle{\sum}^{\lambda\nu}
\end{equation}
\begin{align}
\left\lbrace \textstyle{\sum}^{\mu\nu}, \textstyle{\sum}^{\lambda\sigma}\right\rbrace = 
-\frac{1}{2} & \left(\eta^{\mu\lambda}\eta^{\nu\sigma} -  \eta^{\mu\sigma} \eta^{\nu\lambda}\right) - \frac{i}{2} \epsilon^{\mu\nu\lambda\sigma\rho} \gamma_\rho \\
&(\epsilon^{01235} = +1)\nonumber
\end{align}
\begin{equation}
\left(\textstyle{\sum}^{\mu\nu} L_{\mu\nu}\right)^2 = -\frac{1}{2} L^{\mu\nu} L_{\mu\nu} + 3\textstyle{\sum}^{\mu\nu} L_{\mu\nu}
\end{equation}
\begin{equation}
\delta_{ab}\delta_{cd} = \frac{1}{4} \delta_{ad}\delta_{cb}
+ \frac{1}{4} \gamma_{ad}^\mu \gamma_{\mu cb} - \frac{1}{2} \textstyle{\sum}_{ad}^{\mu\nu} \textstyle{\sum}_{\mu\nu cb}
\end{equation}
\begin{equation}
\textstyle{\sum}_{ab}^{\mu\nu} \delta_{cd} = \frac{1}{4}
 \left( \textstyle{\sum}_{ad}^{\mu\nu}  \delta_{cb} + \delta_{ad} 
\textstyle{\sum}_{cb}^{\mu\nu} \right) + \frac{1}{8}\left( \gamma_{ad}^\nu \gamma_{cb}^\mu - \gamma_{ad}^\mu \gamma_{cb}^\nu \right)
\end{equation}
\begin{equation}
 + \frac{1}{2} \left( \textstyle{\sum}_{ad}^{\mu\lambda} \textstyle{\sum}_{\;\;\;\lambda cb}^{\nu} - 
\textstyle{\sum}_{ad}^{\nu\lambda} \textstyle{\sum}_{\;\;\;\lambda cb}^{\mu}\right) \nonumber 
\end{equation}
\begin{equation}
\hspace{2cm} + \frac{i}{8}\epsilon^{\mu\nu\lambda\sigma\rho} \left( \textstyle{\sum}_{\lambda\sigma ad} \gamma_{\rho cb} + \gamma_{\rho ad} 
\textstyle{\sum}_{\lambda\sigma cb}\right)\nonumber  
 \end{equation}
can be used to show that if
\begin{equation}
A = -\alpha,\qquad B = \frac{1}{2}, \qquad b = 2\alpha^2, \qquad |\!\!\!C = -1, \qquad \gamma = - \alpha
\end{equation}
in eqs. (64) and (65), then the action of eq. (64) is invariant under the
transformations of eq. (65).  Furthermore, the action
\begin{equation}
S^\prime = \int d^4x \delta(x^2 - r^2) \left[ \tilde{\Psi}\Psi + 2 F \Phi - 2 \alpha \Phi^2 \right]
\end{equation}
is also invariant under the transformation of eq. (65) with $\gamma = -\alpha$.

We now can examine the commutator of two transformations of the form given by eq. (65).  If $\delta_i$ is a transformation associated with a parameter $\xi_i$, then it is easily shown that
\begin{subequations}
\begin{align}
\left(\delta_1\delta_2 - \delta_2\delta_1 \right)\Phi &= \left( \tilde{\xi}_2 \textstyle{\sum}^{\mu\nu} \xi_1 - 
\tilde{\xi}_1 \textstyle{\sum}^{\mu\nu} \xi_2\right) L_{\mu\nu}\Phi \nonumber \\
&= 2\tilde{\xi}_2 \textstyle{\sum}^{\mu\nu} \xi_1 L_{\mu\nu} \Phi\\
\left(\delta_1\delta_2 - \delta_2\delta_1 \right)F &= 2 \tilde{\xi}_2 \textstyle{\sum}^{\mu\nu} \xi_1 \left(-3 L_{\mu\nu} \Phi + L_{\mu\nu} F\right)\\
\intertext{and}
\left(\delta_1\delta_2 - \delta_2\delta_1 \right)\Psi &= \tilde{\xi}_2 \textstyle{\sum}^{\mu\nu} \xi_1 (L_{\mu\nu}\Psi) + \frac{i}{2} \epsilon^{\mu\nu\lambda\sigma\rho} (L_{\mu\nu}\gamma_\rho \Psi)  
(\tilde{\xi}_2 \textstyle{\sum}_{\lambda\sigma} \xi_1) .
\end{align}
\end{subequations}
The form of the symmetry transformation given in eqs. (74b,c) is not in accordance with the algebra of eq. (23a).  Consequently we have a model defined on $AdS_4$ which possesses a global Fermionic symmetry whose algebra is not the supersymmetric extension of the algebra of the generators of isometry transformations of this space.

\section{Discussion}
In this paper we have considered a supersymmetric model for the space $AdS_3$ using superspace techniques associated with a four dimensional embedding space.  A two dimensional subspace of this $AdS_3$ space possess a Minkowski space metric and on this space a supersymmetric model has been inferred.

The $AdS_3$ model suggests a supersymmetric model on $AdS_4$; this model has been shown to have a global Fermionic symmetry that does not satisfy the algebra which is the supersymmetric extension of the isometry algebra of $AdS_4$ space.

We hope to extend these considerations to other spaces of constant curvature.  In particular, it would be interesting to find a model on $AdS_5$ that possesses a global Fermionic symmetry and then to use Poincar$\acute{\textrm{e}}$ coordinates on $AdS_5$ to find a model on a four dimensional space with a Minkowski metric that has a novel Fermionic symmetry.

\section*{Acknowledgements}
Conversations with Christian Schubert were helpful. Roger Macleod had useful suggestions.

\end{document}